\documentclass[
reprint,
superscriptaddress,
nobibnotes,
amsmath,
amssymb,
pra,
]{revtex4-2}
\usepackage{graphicx}% Include figure files
\usepackage{dcolumn}% Align table columns on decimal point
\usepackage{bm}% bold math
\usepackage{color}
\usepackage{xcolor}
\usepackage{hyperref}% add hypertext capabilities
\hypersetup{
    colorlinks=true,
    linkcolor=blue,
    filecolor=blue,      
    urlcolor=blue,
    citecolor=blue,
}

\begin{document}

\title{Purcell-enhanced optical refrigeration}

\author{Peng Ju}
\thanks{These authors contributed equally to this work.}
\affiliation{Department of Physics and Astronomy, Purdue University, West Lafayette, Indiana 47907, USA}

\author{Kunhong Shen}
\thanks{These authors contributed equally to this work.}
\affiliation{Department of Physics and Astronomy, Purdue University, West Lafayette, Indiana 47907, USA}

\author{Stefan P{\"{u}}schel}
\affiliation{Leibniz-Institut für Kristallzüchtung (IKZ), Max-Born-Str. 2, Berlin 12489, Germany}

\author{Yuanbin Jin}
\affiliation{Department of Physics and Astronomy, Purdue University, West Lafayette, Indiana 47907, USA}

\author{Hiroki Tanaka}
\affiliation{Leibniz-Institut für Kristallzüchtung (IKZ), Max-Born-Str. 2, Berlin 12489, Germany}

\author{Tongcang Li}
\email{tcli@purdue.edu}
\affiliation{Department of Physics and Astronomy, Purdue University, West Lafayette, Indiana 47907, USA}
\affiliation{Elmore Family School of Electrical and Computer Engineering, Purdue University, West Lafayette, Indiana 47907, USA}
\affiliation{Birck Nanotechnology Center, Purdue University, West Lafayette, Indiana 47907, USA}
\affiliation{Purdue Quantum Science and Engineering Institute, Purdue University, West Lafayette, Indiana 47907, USA}

\date{\today}

\begin{abstract}

Optical refrigeration of solids with anti-Stokes fluorescence has been widely explored as a vibration-free cryogenic cooling technology. A minimum temperature of 87 K has been demonstrated with rare-earth ion doped crystals using optical refrigeration. However, the depletion of the upper-lying energy levels in the ground state manifold hinders further cooling to below the liquid nitrogen (LN$_2$) temperatures, restricting its applications. In this work, we introduce a Purcell-enhanced optical refrigeration method to circumvent this limitation. This approach enhances the emission of high-energy photons by coupling the emitters to an optical cavity, blue shifting the mean emission wavelength. Such Purcell-enhanced emission facilitates cooling starting from a lower energy level in the ground state manifold, which exhibits a higher occupation below the LN$_2$ temperatures. Using experimentally measured optical coefficients, our theoretical analysis predicts a minimum achievable internal temperature of about 38 K for a Yb$^{3+}$:YLiF$_{4}$ nanocrystal near a cavity under realistic conditions. The proposed method is applicable to other rare-earth ion doped materials and semiconductors, and will have applications in creating superconducting and other quantum devices through solid-state cooling.

\end{abstract}

% \keywords{Optical refrigeration, Purcell effect, rare-earth doped crystals, nanocavity}
\maketitle

\section{Introduction}
Optical refrigeration of solids, as a vibration-free cooling method, has made great progress over the last few decades \cite{Richard2007, Mansoor20162}. Since the first demonstration of fluorescence cooling \cite{Epstein1995}, optical refrigeration has been applied to directly cool semiconductors \cite{Qihua2013} and various rare-earth ion doped crystals \cite{Topper:24, Hoyt:03, Rostami:19} as well as nanocrystals \cite{Barker2017}. Efficient fluorescence cooling is also proposed to reduce the temperature of quantum dots \cite{Ricardo2025}. Besides the study of intrinsic cooling, optical refrigeration has been successfully applied to several research domains. Target devices such as mechanical resonators \cite{Peter2020}, quantum defects in diamonds \cite{Dadras:20}, and infrared sensors \cite{Mansoor2018, Mansoor2022} are thermally attached to Yb$^{3+}$ doped crystals and indirectly cooled. Moreover, optical refrigeration can cool both the internal temperature and center-of-mass motion of an optically levitated Yb$^{3+}$ doped nanocrystal \cite{Barker2017, Peter2015, Vamivakas:21, Volz2023}, with the potential for precision measurement \cite{Ahn2020} and studying macroscopic quantum effects \cite{Aspelmeyer2020}. 

The state-of-the-art optical refrigeration is demonstrated with Yb$^{3+}$ doped YLiF$_{4}$ (YLF) crystals, achieving a record low temperature of 87 K \cite{Volpi:19, Mansoor2016}. To maximize cooling efficiency, the Yb$^{3+}$ ions are resonantly pumped from the highest ground-state to the lowest excited-state Stark level. However, the cooling efficiency diminishes due to the population depletion of the absorbing level as the temperature reduces and is eventually dominated by parasitic absorption below 100~K. The current limit on the minimum achievable temperature (MAT) hinders several potential applications of optical refrigeration, such as studying superconductivity \cite{Zaanen2015} and qubits in 2D materials \cite{Liu2019}, which often require temperatures below that of liquid nitrogen (LN$_2$, 77~K at atmospheric pressure). While the parasitic absorption can be mitigated using higher purity host crystals, the implication and practical viability of this approach remains to be investigated \cite{Volpi:19}.  

\begin{figure*}[htp]
\includegraphics[width = 0.98\textwidth]{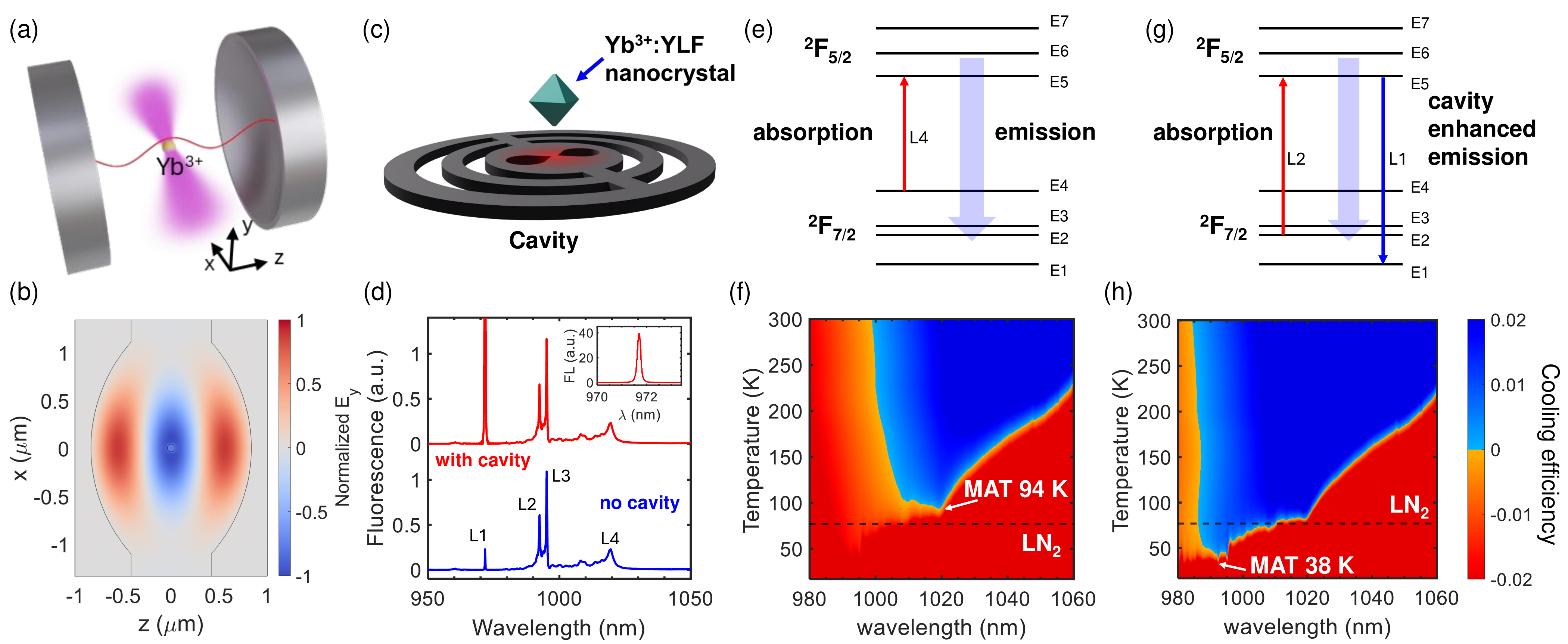}
\caption{\label{fig1} Purcell enhanced optical refrigeration of a Yb$^{3+}$:YLF nanocrystal with an optical cavity. (a) and (c) show two schematic diagrams of the proposed Purcell enhanced optical refrigeration protocol. The emission of high-energy photons from a Yb$^{3+}$:YLF nanocrystal is enhanced by a confocal cavity (a) or a bowtie-shaped nanophotonic cavity (c). The nanocrystal and cavity are separated by a vacuum gap. (b) Finite element method (FEM) simulation of a resonant optical mode of the confocal cavity with a Yb$^{3+}$:YLF nanocrystal levitated at its center, as depicted in the schematic shown in (a). (d) Linearly polarized emission spectrum of Yb$^{3+}$:YLF nanocrystal with (top red) and without (bottom blue) a nearby nanocavity at 50 K. The four labeled peaks of the blue curve correspond to resonant transitions from E5 to E1 (peak 1), E2 (peak 2), E3 (peak 3), and E4 (peak 4), respectively. The inset shows the Purcell enhanced emission peak near 972 nm. (e) Energy diagram of the Yb$^{3+}$:YLF nanocrystal and the conventional optical refrigeration protocol with natural spontaneous emission. Yb$^{3+}$:YLF nanocrystal absorbs 1020 nm photons and emits a broadband fluorescence determined by the branching ratio. (f) Cooling efficiency as a function of temperature and pump wavelength without the cavity. The minimum achievable temperature (MAT) without Purcell enhancement is 94 K when pumped with 1020 nm laser. The horizontal black dashed line indicates the LN$_2$ temperature at 77 K. (g) Energy diagram of Purcell enhanced optical refrigeration of a Yb$^{3+}$:YLF nanocrystal near a nanocavity. The nanocrystal is pumped with 992 nm laser, represented by a solid red arrow. The Purcell enhanced emission is dominated by the resonant transition from E5 to E1, represented as solid blue arrow, in addition to the intrinsic spontaneous emission. (h) Cooling efficiency as a function of temperature and pump wavelength with a cavity that has $F_\text{p}=180$. The MAT for Purcell enhanced optical refrigeration of Yb$^{3+}$:YLF nanocrystal is about 38 K when pumped with 992 nm laser, well below the LN$_2$ temperature.}
\end{figure*}

The emission rate of an emitter can be affected by the electromagnetic density of states at the emission frequency, known as the Purcell effect \cite{Notomi_2010,Andrea2018,Ma2022}. The Purcell factor ($F_\text{p}$) quantifies the enhancement ratio of an emitter over the intrinsic spontaneous emission due to the Purcell effect. For an electric or magnetic dipole resonantly coupled to
a cavity mode, $F_p = \frac{3\lambda^3_n}{4\pi^2}\frac{Q}{V_m}$, where $\lambda_n$ is the wavelength in the material with a refractive index $n$, $Q$ is the quality factor of the mode, and $V_m$ is the effective mode volume. Recently, cavities with high $F_\text{p}$ have been applied to increase the emission rate of rare-earth ions \cite{Tang2023,Hugues2021,Timmerman2020} and atoms \cite{Lukin2013,Lukin2014} near a photonic cavity. Cavities have also been used to cool the motion of atoms trapped in a vacuum \cite{Wolke2012,Lv2023,Lyne2024}. Inspired by these experiments, we propose a protocol to significantly enhance optical refrigeration with cavity-induced Purcell effect and reduce the minimum achievable internal temperature of a Yb$^{3+}$ doped YLF nanocrystal well below the LN$_2$ temperature. In the protocol, the Purcell effect selectively enhances high-energy fluorescence photon emission, causing a blue shift of the mean emission wavelength, thereby enabling the pumping of stronger absorption lines at shorter wavelengths, which remain pertinent for anti-Stokes fluorescence. Based on the experimentally measured absorption and fluorescence spectra, we predict a MAT of about 38 K for a Yb$^{3+}$:YLF nanocrystal near a nanocavity using our proposed optical refrigeration approach. 

\section{Purcell-enhanced cooling scheme}
Two schematic diagrams of our proposed Purcell enhanced optical refrigeration are shown in Fig.~\ref{fig1}(a) and \ref{fig1}(c). A 5$\%$ Yb$^{3+}$:YLF nanocrystal is located inside a Fabry-Perot microcavity \cite{Hugues2021,Benjamin23} or near a nanophotonic cavity \cite{stobbe2022,Kristensen22,Tang2023}, which increases the local density of the photonic states and selectively enhances the emission rate of the nanocrystal around the cavity resonant wavelength by the Purcell effect. The cooling efficiency ($\eta_\text{c}$) of the Yb$^{3+}$:YLF nanocrystal is given by \cite{Richard2007}:
\begin{equation}
    \eta_\text{c}(T, \lambda) = \eta_\text{ext} \eta_\text{abs} (T, \lambda) \frac{\lambda}{\tilde{\lambda}(T)} - 1.
    \label{eq:1}
\end{equation}
Here, $\eta_\text{ext}$ is the external quantum efficiency of Yb$^{3+}$, which takes into account non-radiative decay, the fluorescence trapping, and re-absorption. $\eta_\text{ext}$ is mainly determined by the property of the nanocrystal and we assume it is independent of temperature for simplicity. $\lambda$ is the wavelength of the absorbed photons and $T$ is the temperature of the nanocrystal. $\tilde{\lambda}(T)$ is the mean emission wavelength at $T$. $\tilde{\lambda}(T)$ depends on the Purcell effect which affects emission spectra. $\eta_\text{abs}(T, \lambda) = [1 + \alpha_\text{b}/ \alpha(T, \lambda)]^{-1}$ is the absorption efficiency of the Yb$^{3+}$, where $\alpha_\text{b}$ is the parasitic background absorption and $\alpha(T, \lambda)$ is the absorption coefficient of pumping laser. 

In the proposed method, the cavity tailored to enhance emission at the zero phonon line reduces the mean emission wavelength, thereby increasing cooling efficiency. The Yb$^{3+}$:YLF nanocrystal and photonic cavity are separated by vacuum. Such a design minimizes the thermal contact between them while achieves strong light-matter interaction. Experimentally, this can be achieved by either optical levitation of the nanocrystal \cite{ju2023,Aspelmeyer2018} or bringing a nanocrystal attached to a nanotube \cite{Maruyama2023,Rechnitz2022,Gao2024} near the cavity. For optical levitation, the trapping laser can simultaneously serve as the cooling laser.   

To accurately evaluate the performance of the proposed optical refrigeration method, we measured the intrinsic fluorescence spectra of a bulk Yb$^{3+}$:YLF crystal, with a care to minimize the re-absorption \cite{Hiroki21}, for a temperature range of 20 to 300 K. These can serve as accurate approximations for the fluorescence spectra of Yb$^{3+}$:YLF nanocrystals. The blue curve in Fig.~\ref{fig1}(d) shows the measured linearly polarized fluorescence spectrum at 50~K. The four peaks corresponding to the inter-Stark-level transitions are labeled based on their terminating level according to the energy diagram of Yb$^{3+}$ in YLF shown in Fig.~\ref{fig1}(e). To minimize $\tilde{\lambda}(T)$ and consequently improve cooling efficiency, the cavity is designed to be resonant with the zero-phonon emission peak at 972 nm (Peak 1), which has the highest photon energy among the four major peaks in the emission spectrum. 

The enhancement ratio for the emission near the wavelength of the cavity mode is determined by both the $F_\text{p}$ at the position of the nanocrystal and the linewidth of the cavity. Both the Fabry-Perot microcavities \cite{Hugues2021,Benjamin23} and nanophotonic cavities \cite{stobbe2022,Kristensen22,Tang2023} have been shown to effectively confine photons and enable strong light-matter interactions. A Purcell factor of $F_\text{p}=177$ has been demonstrated for rare-earth ions coupled to a photonic crystal cavity \cite{Tang2023}, while a $F_\text{p}=6000$ was achieved using a topology-optimized bowtie-shaped cavity \cite{stobbe2022}. The geometry and dimensions of these cavities can be fine-tuned to maximize emission enhancement at zero-phonon line. In our calculation, we assume a $F_\text{p}$ of 180 at the location of the Yb$^{3+}$:YLF nanocrystal and a cavity quality factor (Q) of 1100. The above parameters are experimentally feasible according to our simulation with a finite element method (FEM), as shown in Fig.~\ref{fig1}(b) and Appendix \ref{Appendix:B}. The emission spectrum with cavity-induced Purcell enhancement is calculated based on the above parameters, shown as the red curve in Fig.~\ref{fig1}(d). Compared to the spectrum without cavity (blue curve), the mean emission wavelength is blue-shifted from 1004 nm to 982 nm as the cavity selectively enhanced the emission of higher energy (shorter wavelength) photons, shown in Fig.~\ref{Fig:S1}(b) in the appendix. This improves the cooling efficiency of the Yb$^{3+}$:YLF nanocrystal by nearly a factor of 2, assuming the same pump wavelength. In addition to the enhanced cooling power, the blue shift of the mean emission wavelength extends the wavelength for cooling ($\lambda > \tilde{\lambda}$) and allows us to utilize stronger absorption lines to improve pumping rates.

Fig.~\ref{fig1}(e) shows the energy level of Yb$^{3+}$ and the traditional optical refrigeration method (no cavity). The Yb$^{3+}$ ion is pumped with 1020 nm photons, which is the center wavelength of peak 4 shown in Fig.~\ref{fig1}(d). This corresponds to the excitation from E4, the highest energy level of the ground state manifold, to E5, the lowest energy level of the excited state manifold. By resonantly exciting electrons from $^2\text{F}_{7/2}$ to $^2\text{F}_{5/2}$ with the lowest energy photons, the traditional method maximizes $\lambda/\tilde{\lambda}(T)$. This is an ideal protocol when the absorption coefficient ($\alpha$) is much larger than the background absorption coefficient ($\alpha_b$), which is not the case for 1020 nm photons when the internal temperature is below 100 K.   

Without Purcell enhancement, we calculated the cooling efficiency as a function of temperature and the pumping laser wavelength, shown in Fig.~\ref{fig1}(f). The calculation is based on experimentally measured fluorescence spectra of 5$\%$ Yb$^{3+}$:YLF at different temperatures. The details of the calculation are shown in the appendix, where we assume $\eta_{\text{ext}} = 0.996$ and $\alpha_\text{b} = 10^{-4}~\text{cm}^{-1}$ \cite{Mansoor2016}. Without Purcell enhancement, the MAT is about 94~K. This is close to the MAT obtained by state-of-the-art experimental optical refrigeration of Yb$^{3+}$:YLF crystal \cite{Volpi:19, Mansoor2016}.

\begin{figure}[tp]
\includegraphics[width = 0.48\textwidth]{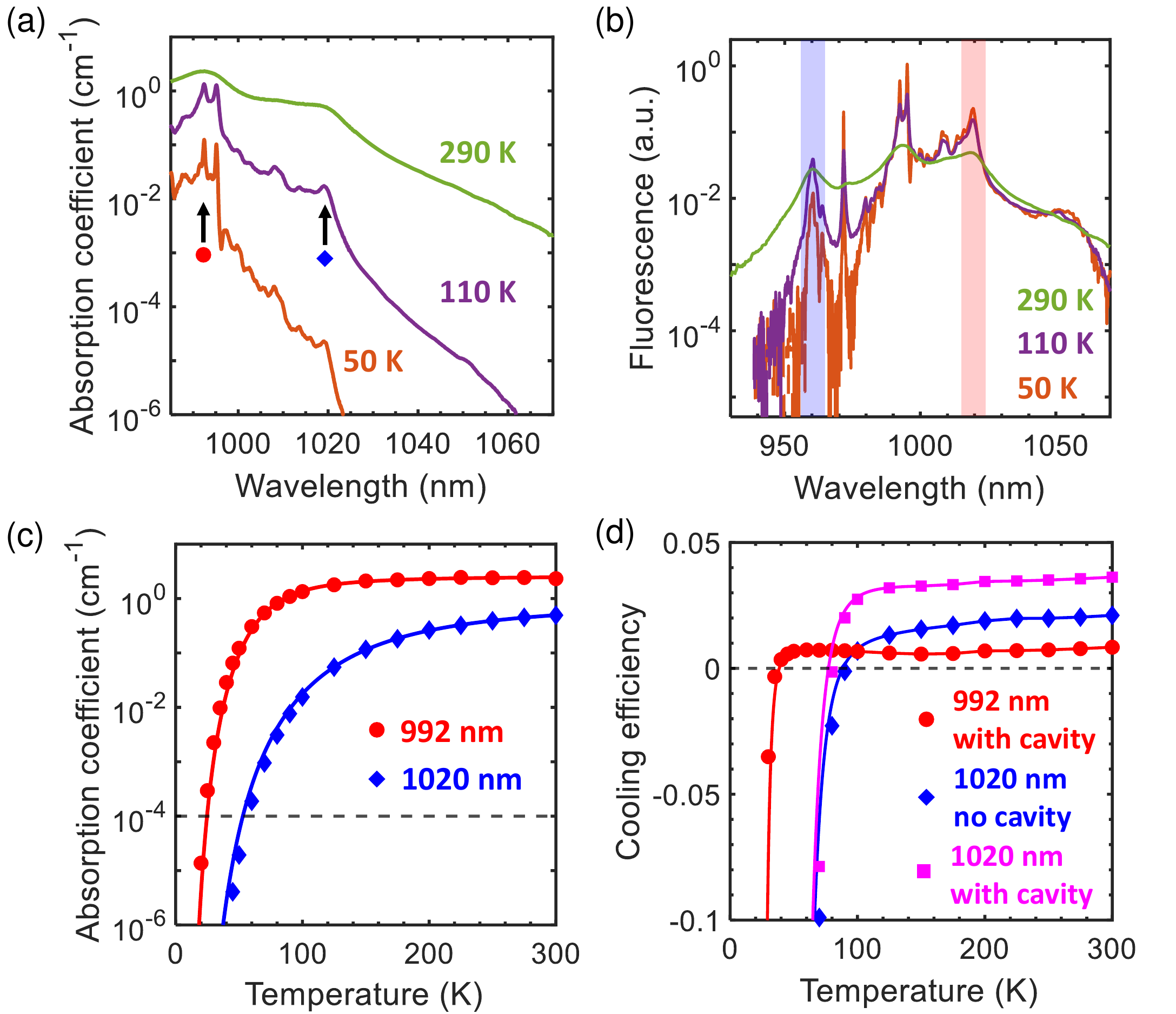}
\caption{\label{fig2} (a) Absorption coefficient of 5$\%$ Yb$^{3+}$:YLF at different temperatures for pi-polarization. The brown, purple, and green curves correspond to the absorption spectra at 50 K, 110 K, and 290 K. The red dot and blue diamond indicate the absorption peaks near 992 nm and 1020 nm. (b) Measured emission spectra of 5\% Yb$^{3+}$:YLF at different temperatures. The brown, purple, and green curves correspond to the measured emission spectra at 50 K, 110 K, and 290 K. The red and blue spectrum ranges indicate the emission bands from E5 to E4 and from E6 to E1, respectively. (c) Measured absorption coefficient of Yb$^{3+}$:YLF as a function of temperature at 992 nm and 1020 nm represented by red dots and blue diamonds, respectively. The red and blue curves correspond to the fitting curves of the measured data. The black dashed line indicates the background absorption coefficient $\alpha_b$. (d) Calculated cooling efficiency as a function of temperature. The red dots show the cooling efficiency for optical pumping at 992 nm with cavity-induced Purcell enhancement. The blue diamonds show the cooling efficiency for optical pumping at 1020 nm without cavity. The magenta squares show cooling efficiency for optical pumping at 1020 nm with cavity-induced Purcell enhancement. The red, blue, and magenta solid curves are the corresponding calculated results based on the fitting curves in (c).}
\end{figure}

In our proposed optical refrigeration protocol (Fig.~\ref{fig1}(g)), the cavity enhances the emission of the zero phonon line at 972 nm via the Purcell effect and blue-shifts the mean emission wavelength (Fig.~\ref{Fig:S1}(b)). This allows us to excite the absorption peak at 992 nm, which shows an absorption coefficient larger than 1 $\text{cm}^{-1}$ at 100~K. The calculated cooling efficiency as a function of temperature and the pumping laser wavelength is shown in Fig.~\ref{fig1}(h). The MAT of the Yb$^{3+}$:YLF crystal is reduced due to the increased absorption coefficient and the blue shift of the mean emission wavelength, as shown in Fig.~\ref{Fig:S1}(a). In particular, the MAT is about 38 K when pumped at the optimum wavelength of 992 nm. The dramatic improvement in MAT with our protocol can potentially enable many applications that require cooling below LN$_2$ temperature \cite{Zaanen2015,Liu2019}. In addition to the significantly improved MAT, optical refrigeration below LN$_2$ temperature can be achieved by absorbing a broad band of photons between 987 nm and 1010 nm, making optical refrigeration a viable substitute for liquid nitrogen cooling systems.

Fig.~\ref{fig2}(a) shows the calculated absorption coefficient spectra based on the reciprocity relation and the measured fluorescence spectra of the 5$\%$ Yb$^{3+}$:YLF crystal at different temperatures, respectively. Both absorption and emission spectra have narrower peaks at lower temperatures due to the reduced thermal crystal phonon energy. The population distribution of the excited state manifold of Yb$^{3+}$:YLF follows the Boltzmann distribution. The red and blue spectrum range in Fig.~\ref{fig2}(b) corresponds to the emission band from E5 to E4 and from E6 to E1, respectively. Therefore, the fluorescence ratio between blue and red bands can be used to characterize the internal temperature of the Yb$^{3+}$:YLF nanocrystal, as shown in Fig.~\ref{Fig:S5}(a).

The higher absorption coefficient at 992 nm than 1020 nm significantly enhances the cooling efficiency of Yb$^{3+}$:YLF nanocrystal. As shown in Fig.~\ref{fig2}(c), the measured absorption coefficient at 992 nm is more than 50 times higher than that at 1020 nm when the temperature is 100 K. This ratio increases at lower temperatures. At a temperature of 50 K, $\eta_\text{abs}$(50 K, 1020 nm) drops to around 0.5 while $\eta_\text{abs}$(50 K, 992 nm) remains nearly unity, which is critical for efficient cooling. The decreased absorption coefficient of 1020 nm is caused by population depletion of E4 state (Fig.~\ref{fig1}(e)) at low temperatures. According to the Boltzmann distribution, the population of E2 state (Fig.~\ref{fig1}(g)) with lower energy is exponentially larger at the same temperature, leading to larger absorption coefficient at 992 nm. Our finding indicates that nanocrystal with lower E2 energy, such as Yb:KY$_3$F$_{10}$ \cite{Hiroki22}, can potentially achieve lower MAT with our proposed optical refrigeration protocol.                                             

Fig.~\ref{fig2}(d) shows the cooling efficiency of different optical refrigeration methods calculated from Eq.~\ref{eq:1}. Without cavity, the cooling efficiency is limited by $\frac{k_B T}{hc/\lambda}$ \cite{Mansoor20162}. Cavity-induced Purcell enhancement can significantly increase the cooling efficiency beyond this limit. Despite enhanced cooling power, the MAT is constrained by the vanishing absorption coefficient of 1020 nm photons at low temperatures, as shown in Fig.~\ref{fig2}(d). In addition to Purcell enhanced emission, our proposed method utilized the optimized pumping wavelength of 992 nm by considering the trade-off between per-photon cooling power and absorption coefficient. The calculated cooling efficiency is nearly constant for temperatures above 50 K, making this an ideal method for optical refrigeration below the LN$_2$ temperature. In practice, both high cooling efficiency at high temperature and low MAT can be achieved by switching pumping laser from 1020 nm to 992 nm at 70 K. Our protocol can easily be adapted to enhance the optical refrigeration of other rare-earth doped particles. 

\begin{figure}[tp]
\includegraphics[width = 0.48\textwidth]{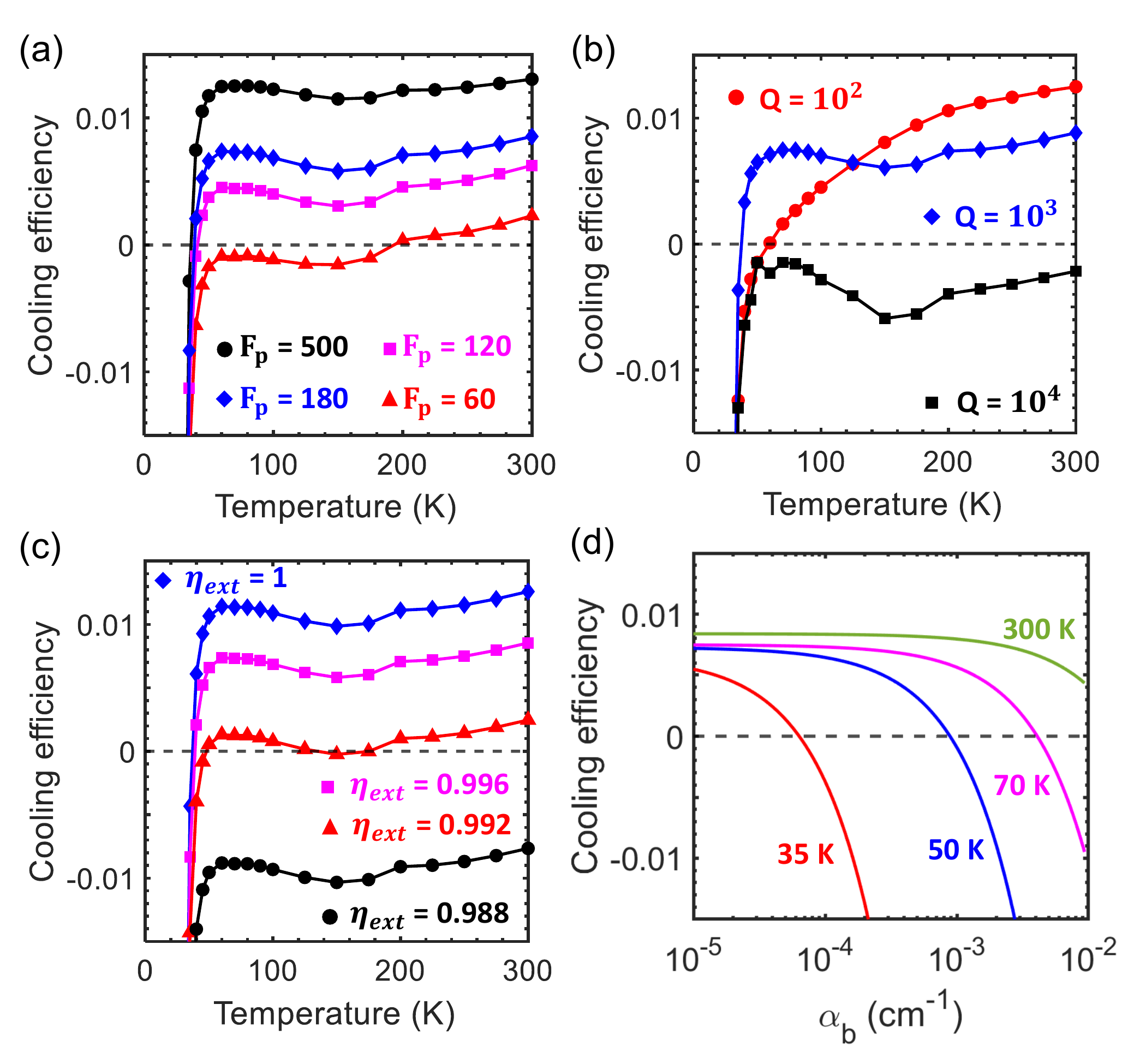}
\caption{\label{fig3} Investigate the impact of experimental conditions on the proposed Purcell enhanced optical refrigeration protocol. (a) Cooling efficiency as a function of temperature for different $F_\text{p}$ as a result of different separation between the Yb$^{3+}$:YLF nanocrystal and the cavity. (b) Cooling efficiency as a function of temperature for different Q factors of the photonic nanocavity. (c) Cooling efficiency as a function of temperature for different external quantum efficiency ($\eta_\text{ext}$) of the Yb$^{3+}$:YLF nanocrystal. (d) Cooling efficiency as a function of the background absorption coefficient ($\alpha_\text{b}$) of the Yb$^{3+}$:YLF nanocrystal for different temperatures.}
\end{figure}

\section{Cooling efficiency}
To study the experimental feasibility of the proposed protocol, we calculated the cooling efficiency of Yb$^{3+}$:YLF nanocrystal in various experimental conditions. Besides the intrinsic properties of Yb$^{3+}$ such as fluorescence, absorption, and energy structure, the properties of cavity and YLF nanocrystal strongly affect the Purcell enhanced cooling efficiency. The $F_\text{p}$ at the location of the nanocrystal and the cavity Q factor collaboratively determines Purcell enhanced emission of high-energy photons, therefore, strongly affects the cooling efficiency. 

The $F_\text{p}$ at the location of nanocrystal depends on the design of the cavity as well as the separation between the nanocrystal and the center of the cavity \cite{Kristensen22, stobbe2022}. Our former calculation is based on a $F_\text{p}$ = 180. As shown in Fig.~\ref{fig3}(a), while increasing the Purcell factor beyond $F_\text{p}$ = 120 enhances the cooling efficiency, this change has a limited effect on the MAT. The Q factor of the cavity determines the wavelength range enhanced by the Purcell effect. To optimize MAT, the linewidth of the cavity should match and cover the linewidth of the zero-phonon emission peak. For this reason, Q = 1000 leads to a better cooling effect than larger or smaller Q, as shown in Fig.~\ref{fig3}(b). 

In Fig.~\ref{fig3}(c), we calculated the cooling efficiency as a function of temperature for different values of external quantum efficiency. For each curve, the external quantum efficiency was assumed to be constant, which represents a simplification. If the external quantum efficiency varies with temperature, the corresponding cooling efficiency can be obtained from the appropriate point on the curves for the given temperature and efficiency. Furthermore, the external quantum efficiency is generally expected to increase as the temperature decreases. This trend is supported by the reported external quantum efficiency values for Tm:ZBLANP \cite{Hoyt2003} and by the measured temperature-dependent fluorescence lifetimes of Yb:YLF \cite{Hiroki21}.

Improving the purity of the material has been the focus for achieving cryogenic cooling with Yb$^{3+}$:YLF crystals. In the past few decades, great progress has made to synthesize high purity Yb$^{3+}$:YLF with near unity quantum efficiency and a remarkably low background absorption coefficient of about $10^{-4}~ \text{cm}^{-1}$ \cite{Mansoor2016}. From Fig.~\ref{fig3}(d), the cooling efficiency at 50 K remains positive for $\alpha_b$ up to 8.7$\times 10^{-4}~ \text{cm}^{-1}$. Although further improved $\alpha_\text{b}$ is required to achieve positive cooling below 38 K, the above analysis shows that our proposed Purcell-enhanced optical refrigeration method is capable of achieving cooling below the LN$_2$ temperature under realistic experimental conditions.

\section{Summary}
In the paper, we proposed a novel Purcell enhanced optical refrigeration protocol. For a Yb$^{3+}$:YLF nanocrystal levitated in a cavity with $F_\text{p}=180$, the minimum achievable temperature is predicted to be about 38 K based on measured fluorescence spectra. Our study of the protocol is conducted over various experimental conditions of cavity and nanocrystal, validating the feasibility of the method. Our protocol paves the way for various applications of optical refrigeration below LN$_2$ temperatures and can be adopted to improve the MAT of several solid-state coolers \cite{Hiroki22, Vamivakas:21}. This scheme will be particularly useful for cooling the internal temperature of optomechanical systems containing spin defects \cite{Dadras:20,Jin2024,Gao2025}.

\vspace{12pt}
\begin{acknowledgments}
We thank Chen-Lung Hung, Xinchao Zhou, Peter J Pauzauskie for helpful discussions. We acknowledge the support from the National Science Foundation under Grants PHY-2409607 and EEC-2224960 (Center for Quantum Technologies \cite{Stewart2026}), and the Office of Naval Research under Grant No. N00014-18-1-2371.
\end{acknowledgments}

\section*{DATA AVAILABILITY}
The data that support the findings of this study are openly available at \cite{data}.

\appendix

\section{Purcell enhanced emission}

For a dipole coupled to an electromagnetic resonator, the probability of spontaneous emission is increased by a factor, called the Purcell factor\cite{purcell1946}:

\begin{equation}
    F_p = \frac{3\lambda^3_n}{4\pi^2} \frac{Q}{V_{m}}.
\end{equation}

Here, $V_m$ is the effective mode volume, corresponding to the effective volume of a single photon. $\lambda_n = \lambda/n$ is the wavelength in the material with the refractive index n. The calculated $F_p$ is the Purcell factor at the position of the maximum electric field. A more general equation of $F_p$ considering the electric field distribution is:
\begin{equation}
    F_p(r) = \frac{3\lambda^3_n}{4\pi^2} \frac{Q}{V_{m}} \frac{|E(r)|^2}{|E_\text{max}|^2} .
\end{equation}
This indicates that the Purcell enhancement depends on the dipole's location \cite{Tang2023}. In our proposed protocol, we consider a confocal microcavity or a topology-optimized bowtie-shaped nanocavity, which has demonstrated a Purcell factor of 6 $\times$ 10$^3$ over the bandwidth of 2 nm at the center of the cavity \cite{stobbe2022, Kristensen22}. The main text considers a Yb$^{3+}$:YLF nanocrystal located at sub-micrometer away from the cavity, where the Purcell factor is about 180. The Purcell factor increases as the nanocrystal approaches the surface of the nanocavity. 

\begin{figure}[tp]
    \centering
    \includegraphics[width = 0.48\textwidth]{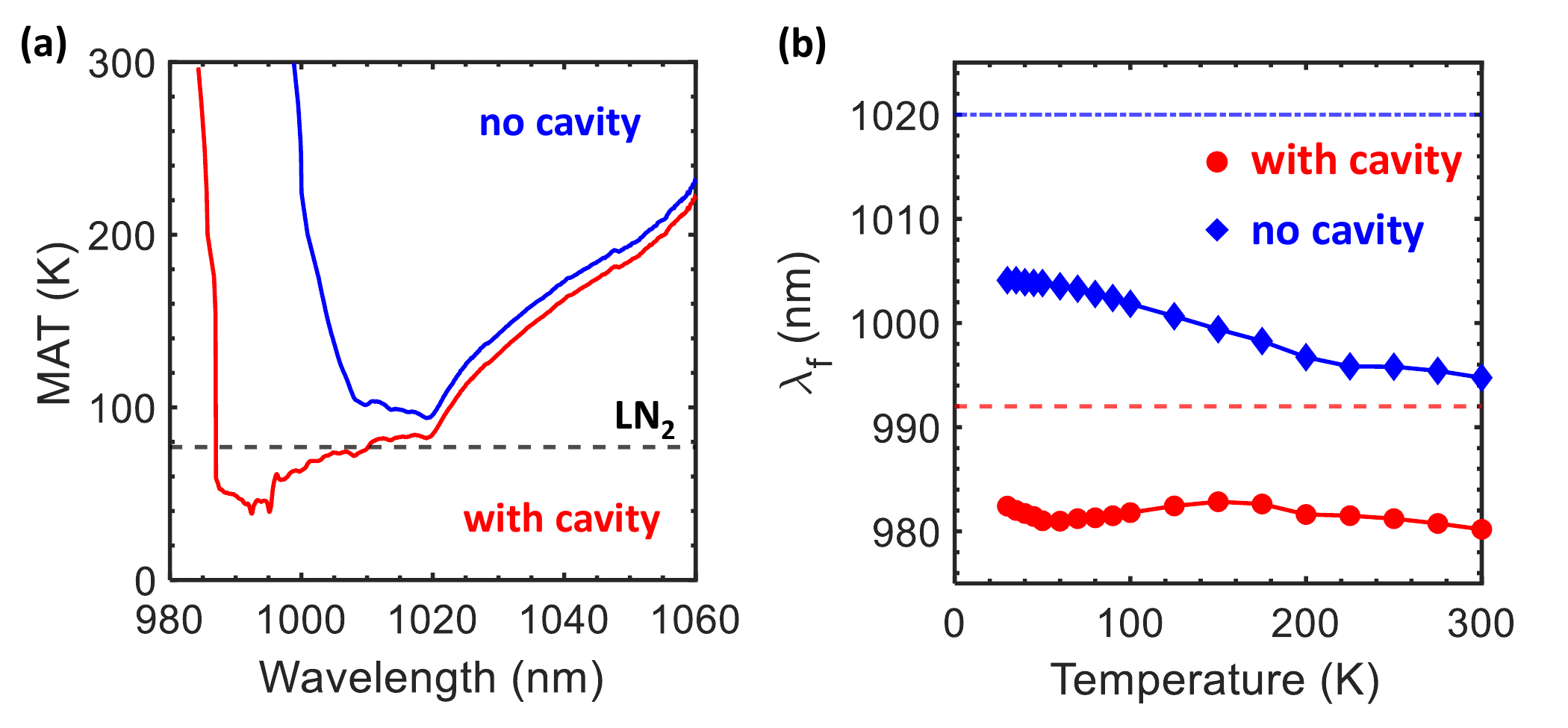}
    \caption{(a) Minimum achievable temperature (MAT) as a function of absorption wavelength. The blue curve corresponds to the MAT of the conventional optical refrigeration protocol without a cavity. The Yb$^{3+}$:YLF nanoparticle absorbs 1020 nm photons. The red curve corresponds to the MAT of our proposed protocol. The Yb$^{3+}$:YLF nanoparticle absorbs 992 nm photons, and the emission of high-energy photons is enhanced by the Purcell effect of a cavity. The black dashed line indicates the temperature of liquid nitrogen. (b) Calculated mean emission wavelength of different protocols. The red dots are the mean emission wavelength with Purcell enhancement. The red dashed line indicates the photon energy of the exciting laser at 992 nm. The blue diamonds are the mean emission wavelength without Purcell enhancement. The blue dashed line indicates the photon energy of the exciting laser at 1020 nm.}
    \label{Fig:S1}
\end{figure}

The Purcell factor $F_p$ represents the enhancement factor for spontaneous emission at the resonant wavelength of the cavity $\lambda_c$. As the emission peak has a finite linewidth, we need to consider the Purcell factor at near resonant emission wavelength given by:
\begin{equation}
    P(\lambda) = P(\lambda_c) \times \frac{(\lambda_c/Q)^2}{(\lambda - \lambda_c)^2  + (\lambda_c/Q)^2},
\end{equation}
where $P(\lambda_c)$ is the Purcell factor at the cavity resonant wavelength. The emission spectrum is engineered by the wavelength-relevant Purcell factor. The resulting emission spectrum near the resonant wavelength can be calculated by convolving the original emission spectrum with the wavelength-dependent Purcell factor. Therefore, to maximize the Purcell enhanced emission near the resonant wavelength at low temperature, the linewidth of the cavity should match and cover the linewidth of the zero-phonon emission peak. The effective quality factor of the emission peak near 972 nm is close to 500 at 50 K. As a result, a cavity with Q = 1000 produces a better cooling effect than those with Q = 100 and Q = 10000, shown in the main text Fig.~3(b).  

The Purcell enhanced emission of high-energy photons shifts the averaged fluorescence wavelength, as shown in Fig.~\ref{Fig:S1}(b). When there is no cavity-induced Purcell enhancement, the averaged fluorescence wavelength increases as the nanocrystal temperature decreases. This reduces the per-photon cooling power at low temperatures, preventing cooling below liquid nitrogen temperatures. In the absence of Purcell enhanced emission, the averaged fluorescence wavelength is changing from 994 nm to 1004 nm, which has higher energy than photons of 992 nm. Therefore, optical refrigeration employing the high absorption coefficient of 992 nm photons cannot be achieved with the conventional protocol.

In our proposed optical refrigeration protocol, the emission of high energy photons is selectively enhanced, blue-shifting the average fluorescence wavelength, shown as the red curve in Fig.~\ref{Fig:S1}(b). This increases the per-photon cooling power of the Yb$^{3+}$:YLF nanocrystal compared to the conventional protocol, reducing the Minimum Achievable Temperature (MAT) as shown in Fig.~\ref{Fig:S1}(a). In particular, the blue-shifting of the average fluorescence enables efficient cooling with the absorption of 992 nm photons, which corresponds to the resonant absorption of zero-phonon line. Due to the high absorption coefficient of 992 nm photons at low temperature, our proposed optical refrigeration protocol remains effective below liquid nitrogen temperatures. The MAT of our protocol reaches 38 K when the absorption photon wavelength is around 992 nm. 

\section{FEM Simulation of the cavity mode and the Purcell factor} \label{Appendix:B}
As we mentioned in the main text of the manuscript, our proposed Purcell enhanced optical refrigeration method can be implemented in multiple systems: (1) a Yb$^{3+}$:YLF nanoparticle optically levitated in a confocal microcavity and (2) a suspended Yb$^{3+}$:YLF nanoparticle near a bowtie-shaped cavity. Here we conduct finite element (FEM) simulation of the Purcell effect between Yb$^{3+}$:YLF nanoparticle and the cavity in both systems.

\subsection{Confocal microcavity}
\begin{figure}[tp]
    \centering
    \includegraphics[width=0.48\textwidth]{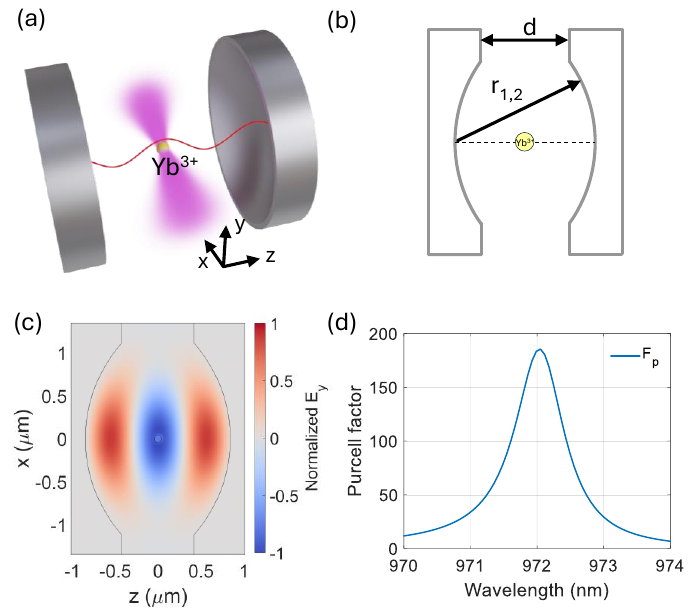}
    \caption{FEM simulation of a Yb$^{3+}$:YLF nanoparticle in the confocal cavity. (a) Schematic of a nanoparticle ($\mathrm{Yb^{3+}:YLF}$) optically levitated at the center of a micro-cavity. The confocal design strongly confines the photons and enhances the spontaneous emission of the Yb$^{3+}$:YLF nanoparticle. (b) The cross-section of the micro-cavity with a confocal design. $r_{1,2}$ is the radius of the respective concave mirror. Two mirrors are separated by a gap $d$. (c) Normalized $z$-component of the electric field in the cross-section of the micro-cavity with a dipole excitation inside the nanoparticle. Three anti-nodes of the standing wave are seen in the simulation results. (d) The response of the micro-cavity as a function of excitation wavelength. The resonance peak is located at $\lambda = 972$ nm with a Purcell factor of 188.}
    \label{fig:confocal_cavity}
\end{figure}

Starting from the definition of the Purcell factor for a dipole emitter, we build an electromagnetic model of a Yb$^{3+}$:YLF nanoparticle levitated in a confocal microcavity. We conducted an FEM simulation with COMSOL to compute the quality factor (Q) and the Purcell factor ($F_p$) of the microcavity. Fig.~\ref{fig:confocal_cavity}(a) is the schematic of a Yb$^{3+}$:YLF nanoparticle levitated in a microcavity. The parameters of the microcavity can be found in Fig.~\ref{fig:confocal_cavity}(b), where the radii of the two cavity mirrors are $r_1=r_2=r_m=1680$ nm. Both cavity mirrors share the same focus point, known as a confocal cavity. Here, the parameter $d$ defines the opening size of the micro-cavity. In the simulation, we set $d=r_m/2$ to provide a reserved window for optical levitation of a Yb$^{3+}$:YLF nanoparticle at the center of the micro-cavity using a separate laser propagating in the direction perpendicular to the axis of the microcavity. The radius $r_\text{sp}$ of the Yb$^{3+}$:YLF nanoparticle is set to 40 nm.

A general formula of the power reflectivity for normal incident light is: 
\begin{equation}
    \text{Reflectivity}=\left\vert \frac{\tilde{n}-1}{\tilde{n}+1} \right\vert^2 = \frac{(n-1)^2+k^2}{(n+1)^2+k^2},
    \label{eq:reflectivity}
\end{equation}
where $\tilde{n}=n+ik$ is the complex refractive index of the material (whether metal, dielectric, or metamaterial). 

The wavelength of the transition E5$\rightarrow$E1 is 972 nm, corresponding to a reflectivity of 0.984 for a gold-coated mirror. Here, the complex refractive index of gold at 972~nm is $\tilde{n}=0.17033+6.4229i$ \cite{Magnozzi2019}. With multi-layer thin  dielectric material coating technique, the reflectivity of a mirror can be improved to be more than 0.999. In the simulation, $k=24$ is used to imitate a mirror with an reflectivity of 0.9988. We perform the eigen-frequency study using FEM simulation in COMSOL to find the eigen-modes of the cavity, which have a Q factor around 1,000.

To calculate the Purcell factor of this micro-cavity, we add an oscillating electric dipole source ($\vec{p}=[0,1,0]$ A/m, where we use the surface current density as the boundary condition in the COMSOL simulation) at the origin point. A spherical boundary is added to cover the dipole source and calculate the radiation power. We compare the radiation power with and without the micro-cavity to compute the Purcell factor \cite{Xu2023Nanocavity}:
\begin{equation}
    F_p = P_\text{rad}^\text{w/}/P_\text{rad}^\text{w/o}.
\end{equation}
A simulated Purcell factor of 188 at $\lambda = 972$ nm is obtained using the above parameters. The dimension of the micro-cavity can be adjusted to selectively enhance the emission peaks shown in Fig.~1(d) of the main text. 

\begin{figure}[tp]
    \centering
    \includegraphics[width=0.48\textwidth]{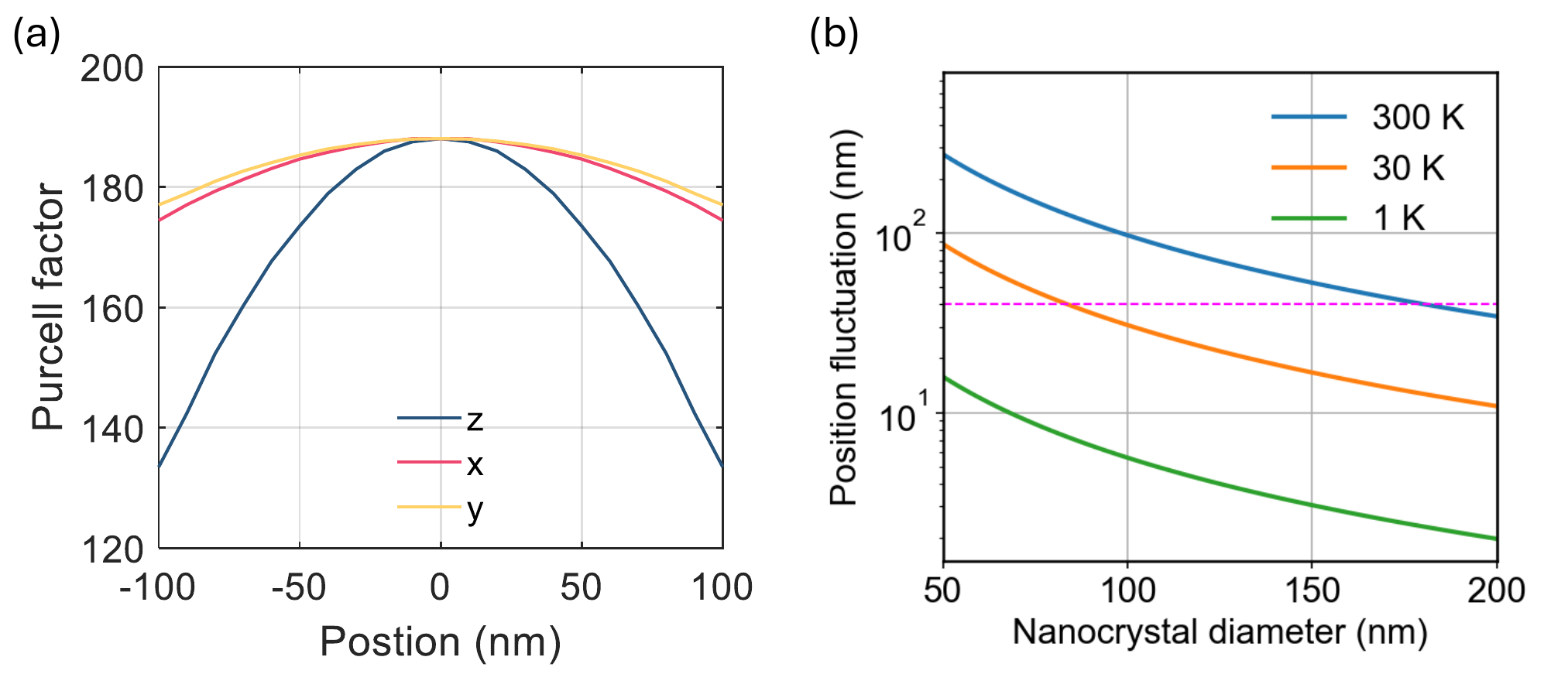}
    \caption{(a) Purcell factor as a function of COM displacement for a nanoparticle levitated near the center of the micro-cavity. (b) Amplitude of position fluctuations of a levitated nanoparticle due to Brownian motion at different temperatures. The trapping frequency is assumed to be 100 kHz. The horizontal dashed line marks 40 nm.}
    \label{fig:Fp_shift}
\end{figure}

In addition to the simplified schematic where a nanoparticle is fixed at the center of the micro-cavity, we simulate the Purcell factor as a function of COM displacement due to thermal Brownian motion in realistic optical levitation experiments. From the simulated results shown in Fig.~\ref{fig:Fp_shift}(a), the Purcell factor is more sensitive to displacement along the $z$ axis (along the cavity axis) compared to the $x$ and $y$ axes. The Purcell factor is greater than 180 for $|z| < 40~nm$. Assuming a trapping frequency of 100 kHz, this can be achieved by a moderate feedback cooling of the COM of the levitated nanoparticle to below 96 K. This is doable as the COM of a levitated nanoparticle has been cooled to below 1 K before \cite{bang2020five}.

\subsection{Bowtie cavity}

\begin{figure*}[tp]
    \centering
    \includegraphics[width=0.9\textwidth]{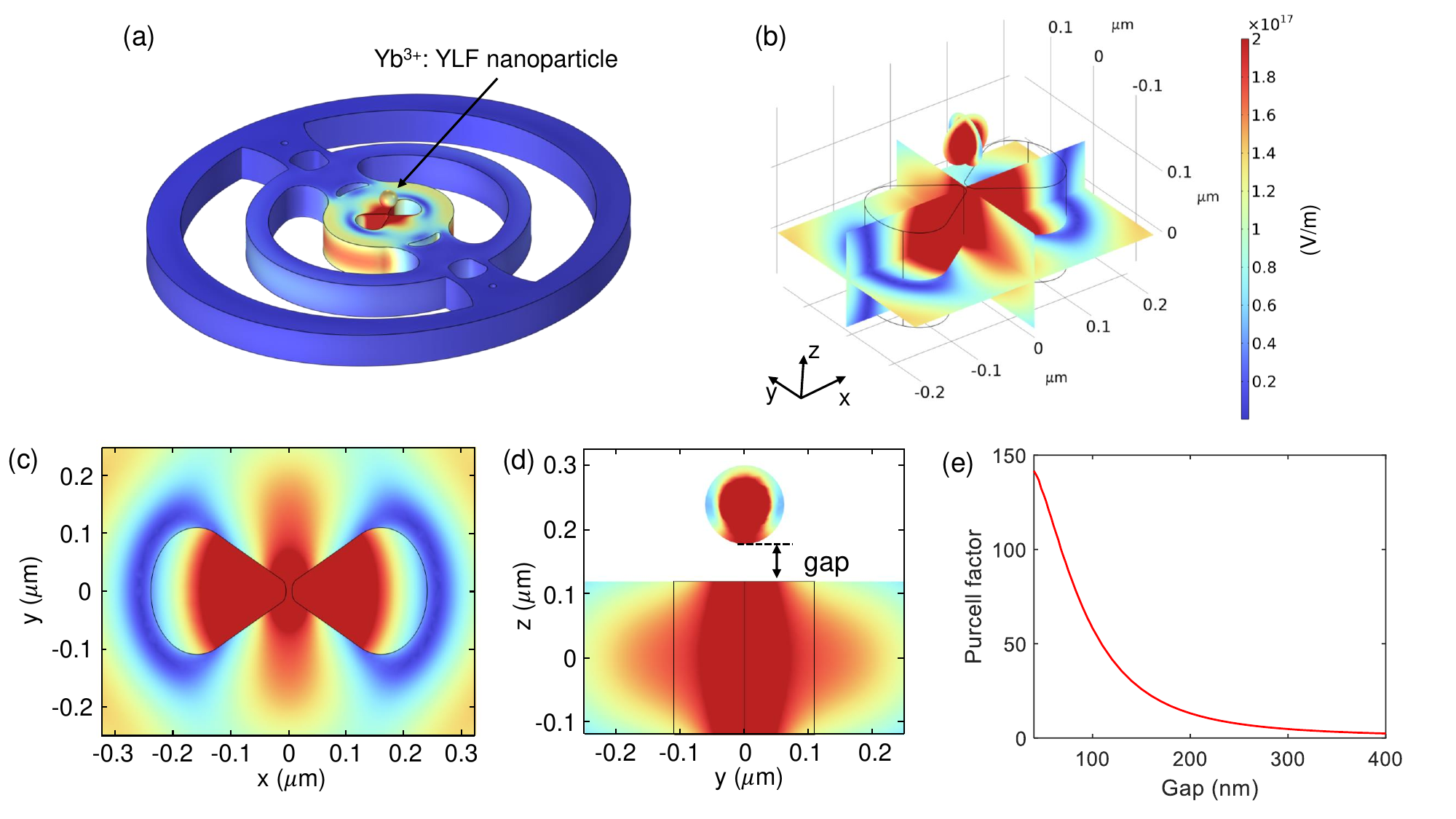}
    \caption{FEM simulation of a $\mathrm{Yb^{3+}:YLF}$ nanoparticle near a Bowtie cavity. (a) schematic of a nanoparticle ($\mathrm{Yb^{3+}:YLF}$) suspended above the center of the bowtie cavity by a nanotube. The nanoparticle is placed a few tens of nanometers away from the cavity surface. The surface color shows the amplitude of the E-field when we excite the cavity with an electric dipole source placed in the center of the nanosphere. (b), (c) and (d) present the E-field of the bowtie cavity from 3D view, $xy$ plane, and $yz$ plane, respectively.  (a)-(d) subfigures share the same colormap displayed in (b). The bowtie structure greatly squeezes the electric field to a tiny mode volume ($\ll\lambda^3$). (e) Purcell factor at the center of the nanoparticle as a function of the gap between the nanosphere and the bowtie cavity. The Purcell factor drops exponentially as the gap increases.}
    \label{fig:bowtie_cavity}
\end{figure*}

Using a similar method, we also simulated the electric field of a Yb$^{3+}$:YLF nanoparticle near a bowtie cavity using COMSOL. The geometry of the bowtie cavity, shown in Fig.~\ref{fig:bowtie_cavity}(a), is inspired by a previous research work \cite{Kristensen22}. The bowtie cavity is made of gallium phosphide (GaP) with a thickness of 155 nm. A nanosphere ($r_\text{sp}=40$ nm) embedded with an electric dipole source is placed above the center of the bowtie cavity. Based on a unit oscillating dipole amplitude (1 A/m) along the $y$ axis, we integrate the radiation power of the dipole source with(w/) and without(w/o) the cavity to calculate the Purcell factor at the position of the nanoparticle.

% build a bowtie cavity and analyze its stimulated electric field by COMSOL. The geometry of the bowtie cavity (shown in Figure.~S\ref{fig:bowtie_cavity}(a)) is inspired from the previous research work \cite{Kountouris22bowtie}. The bowtie cavity is made of Gallium Phosphide (GaP) with a thickness of 155 nm. We place a nanosphere ($r_\text{sp}=40$ nm) embedded with an electric dipole source above the cavity. The dipole amplitude is 1 A/m along the $y$ axis. Like the micro-cavity, we can integrate the radiation power of the dipole source with/without the cavity to calculate the Purcell factor at a certain point.

The results of the FEM simulation are shown in Fig.~\ref{fig:bowtie_cavity}. For Fig.~\ref{fig:bowtie_cavity}(a-d), we assume the nanoparticle is suspended 80 nm above the center of the bowtie cavity, with a surface-to-surface gap of 40 nm. The electromagnetic fields are strongly confined in the bowtie cavity, creating a strong Purcell enhancement for emitters near the cavity. Fig.~\ref{fig:bowtie_cavity}(e) shows the Purcell factor as a function of the gap between the nanosphere and the bowtie cavity. It drops as the gap increases. At gap = 40 nm, the Purcell factor is equal to 145. The nanoparticle may be attached to a boron nitride nanotube \cite{Gao2024} to be positioned at such a small separation from the bowtie cavity.

\section{Temperature dependent fluorescence}

Here we demonstrate that the fluorescence ratio between two bands of Yb$^{3+}$:YLF emission spectrum can be used to measure the internal temperature of the nanocrystal. At the current temperature regime, the lifetime of a Yb$^{3+}$ excited state is about 2 ms \cite{Hiroki21}, far longer than the thermalization of the ground/excited state manifold. So we can assume that the occupation probability of the ground/excited state manifold simply follows the Boltzmann distribution. The energy diagram of Yb$^{3+}$ is shown in Fig.~1(e) in the main text, where the ground states have four energy levels: E1, E2, E3, and E4. The occupation probability of the ground state manifold is given by: 
\begin{equation}
    p_i = \exp(-E_i/k_BT)/\sum^{4}_{j=1}{\exp(-E_j/k_BT)},
\end{equation}
where i = 1, 2, 3, and 4 stand for the ground energy level of Yb$^{3+}$. 

\begin{figure}[tp]
\includegraphics[width = 0.48\textwidth]{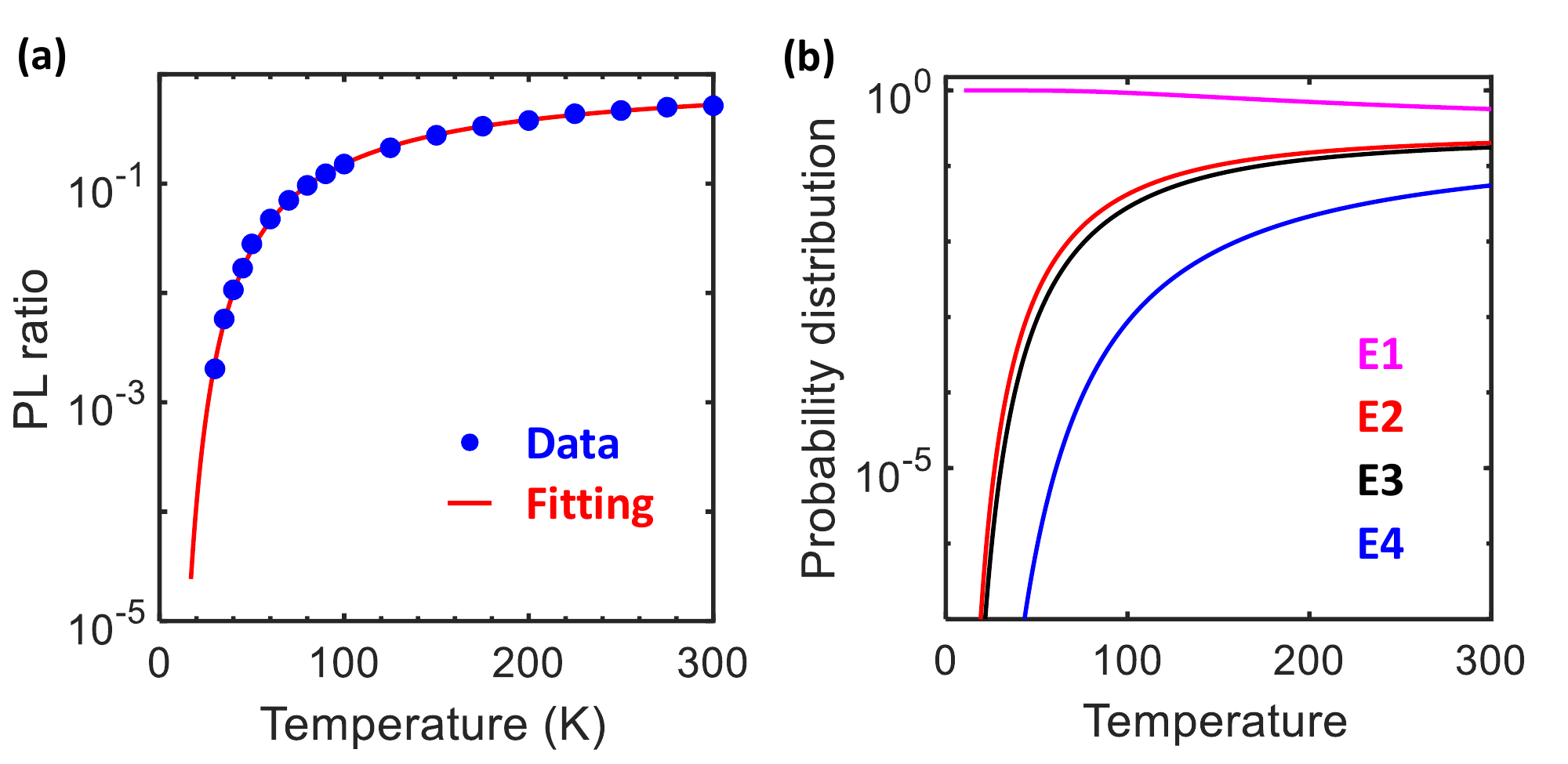}
\caption{\label{Fig:S5} (a) Fluorescence ratio between the 960 nm and 1020 nm as a function of temperature. The blue circles are calculated fluorescence ratios between E5-E4 and E6-E1 from the measured data. The data is fitted with the Boltzmann distribution, shown as the red curve. (b) Calculated ground state probability distribution of the Yb$^{3+}$:YLF nanocrystal as a function of temperature. The ground state energy level is shown in the main text Fig.~1(e). The cyan, red, black, and blue curves are the probability distribution of E1, E2, E3, and E4, respectively. }
\end{figure}

Fig.~\ref{Fig:S5}(b) shows the occupation probability as a function of temperature for the ground state manifold. As the temperature of the crystal decreases, the occupation probability of the E1 state increases to unity, while the occupation probability of the E2, E3, and E4 states rapidly decreases. This leads to a vanishing absorption coefficient near peak 2{\textendash}5 at low temperatures, diminishing cooling efficiency. At T = 50 K, $p_4/p_2 = 2268$, and the absorption coefficient at 992 nm is 6700 times higher than that at 1020 nm.

The occupation probability of the excited state manifold also follows the Boltzmann distribution:
\begin{equation}
    p_i = \exp(-E_i/k_BT)/\sum^{7}_{j=5}{\exp(-E_j/k_BT)},
\end{equation}
where, i = 5, 6, and 7 refer to the excited energy levels of Yb$^{3+}$. The intrinsic spontaneous emission rate from the excited state to the ground state depends on the occupation probability of the excited state and the branch ratio. As the branching ratios are independent of temperature, the temperature dependence of spontaneous emission is dominated by the excited state occupation. We calculated the fluorescence ratio of the transition band E5-E4 and band E6-E1 from the experimentally measured emission spectrum, shown in Fig.~\ref{fig2}(b) in the main text. The data fit well with the equation derived from the Boltzmann distribution. In the proposed optical refrigeration protocol, the emission band near 992 nm is increased to enhance the cooling efficiency. The emission bands near 960 nm and 1020 nm can be employed to measure the internal temperature of the nanocrystal.  

\section{Effect of background absorption on cooling efficiency}

\begin{figure}[thp]
\includegraphics[width = 0.3\textwidth]{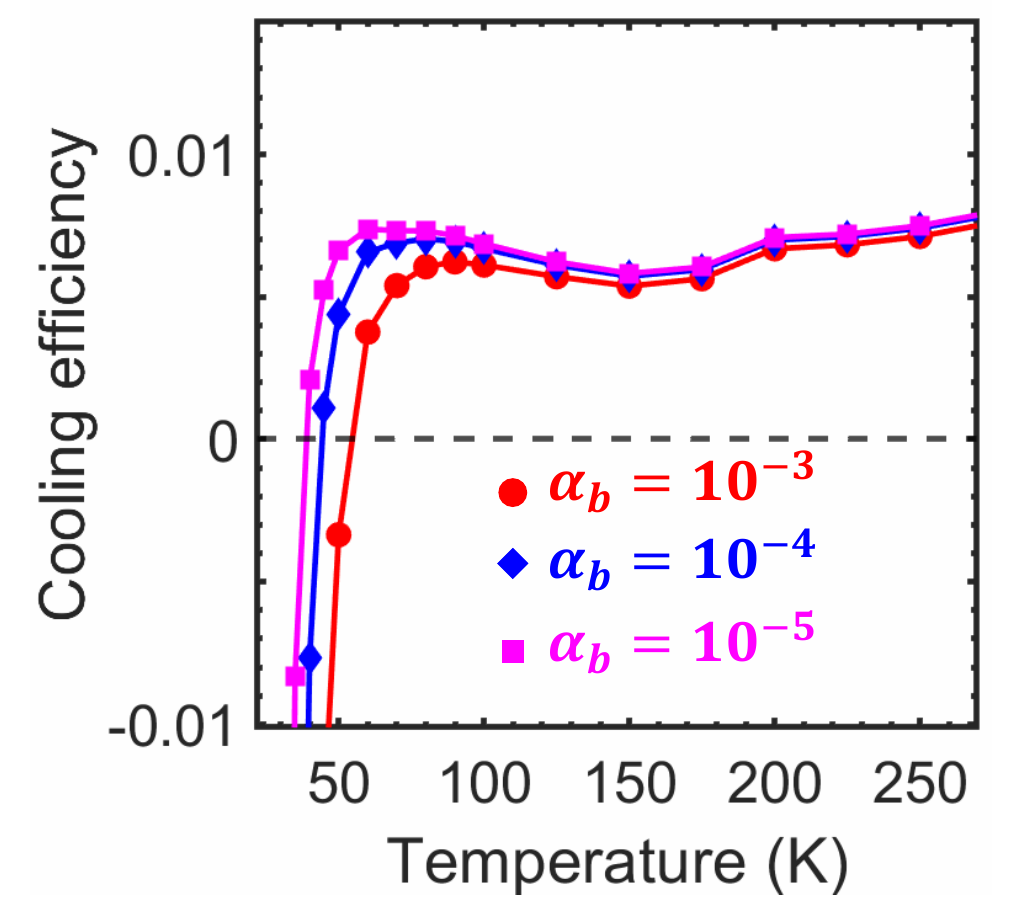}
\caption{\label{Fig:S6} Cooling efficiency as a function of temperature for different background absorption coefficients $\alpha_b$. The red dots, blue diamonds, and cyan squares correspond to the cooling efficiency of $\alpha_b$ = 10$^{-3}$, 10$^{-4}$, and 10$^{-5}$. The impact of $\alpha_b$ is small when the temperature is above 100 K. }
\end{figure}

Based on our proposed optical refrigeration protocol, we calculate the cooling efficiency as a function of temperature at different background absorption coefficients $\alpha_b$, as shown in Fig.~\ref{Fig:S6}. The cooling efficiency is nearly not affected by changing the $\alpha_b$ when the temperature is above 100 K. Below 100 K, the cooling efficiency for high background absorption (red dot) decreases to negative around 50 K, while the cooling efficiency for low background absorption (pink square) remains high even at 35 K. This indicates that while reducing background absorption does not have an effect on cooling above 100 K, it is critical for achieving a MAT lower than 50 K. 

In the ultimate limit where the crystal exhibits no background absorption, the doped rare-earth ions would behave much like trapped ions in ultrahigh vacuum. In this regime, the MAT could fall below 1 K, since millikelvin and even microkelvin cooling has been demonstrated with trapped ions in vacuum. Here, background absorption plays a role analogous to that of residual air molecules for trapped ions. At sufficiently low temperatures, other factors, such as the vibrational motion, will ultimately limit the MAT. 

Based on Ref. \cite{Tang2023}, the decay rate of Er ions in a cavity with a Purcell factor of 177 is 71.4~kHz. The decay rate of Yb$^{3+}$ ions is expected to be of the same order, and we assume a cavity-enhanced decay rate of 50~kHz. Each optical pumping and decay cycle reduces the energy by $\Delta E \approx$ 26~meV. 
For an 80~nm YLF nanoparticle doped with 5\% Yb$^{3+}$ ions, the maximum net cooling power can be estimated to be about 38 pW. If the cooling efficiency under certain laser excitation is 0.01, the corresponding absorbed laser power is 3.80~nW, yielding a laser cooling power of 3.84~nW and a net cooling power of 38~pW. Here the cooling efficiency $\eta_c=$ (net cooling power) / (absorbed laser power), with (net cooling power) = (laser cooling power) – (laser heating power).  For comparison, laser heating in common levitated SiO$_2$ nanoparticles in optical tweezers depends on the optical absorption coefficient and the tweezer power. Assuming an optical absorption coefficient of 10 dB/km (about 100 times worse than that of a high-quality optical fiber) and a tweezer power of 0.1 W with a 1 $\mu$m waist, the heating on an 80-nm SiO$_2$ nanoparticle is about 0.1 pW. Thus, the estimate net cooling power of a Yb$^{3+}$ doped nanocrystal ($\sim$38 pW) is much larger than the heating power ($\sim$0.1 pW) of a typical optically levitated SiO$_2$ nanoparticle.

\end{document}